\documentclass[10pt,twocolumn,cleanfoot]{config/asme2e}

\usepackage{empheq, amssymb}

\usepackage{lmodern}
\usepackage{amsmath}
\usepackage{amsfonts}
\usepackage{amssymb}

\usepackage{bm} 

\usepackage{txfonts}

\usepackage[T1]{fontenc}
\usepackage[utf8]{inputenc} 

\usepackage[dvipsnames]{xcolor} 

\usepackage[american]{babel} 

\usepackage{graphicx}

\usepackage{epstopdf}

\usepackage[noadjust]{cite} 

\usepackage{ltxcmds}

\usepackage{mathtools}

\usepackage{subcaption}

\usepackage{multirow} 

\usepackage{blindtext}

\usepackage{hyperref}

\usepackage{enumitem}

\usepackage{ifthen}

\usepackage{etoolbox}

\usepackage{cancel}

\usepackage{empheq}

\usepackage{fontawesome}

\usepackage{soul}

\usepackage{balance}

\usepackage{textcomp}

\usepackage{lipsum}

\definecolor{light-gray}{gray}{0.4}
\definecolor{box-gray}{gray}{1}

\hypersetup{
    unicode=false,          
    pdftoolbar=true,        
    pdfmenubar=true,        
    pdffitwindow=false,     
    pdfstartview={FitV},    
    pdftitle={Design-OS: A Specification-Driven Framework for Engineering System Design with a Control-Systems Design Case},    
    pdfauthor={H. Sinan Bank and Daniel R. Herber and Thomas H. Bradley},     
    pdfsubject={ASME IDETC/CIE 2026},   
    pdfkeywords = {Design Theory and Methodology, Design Process, Systems Engineering, AI/KBS, Control, Mechatronics and Electro-Mechanical Systems, Design Automation}, 
    pdfnewwindow=true,      
    colorlinks=true,
    allcolors=blue
}

\usepackage{nomencl}

\renewcommand\nomgroup[1]{%
  \item[\bfseries
  \ifstrequal{#1}{V}{ Variables}{%
  \ifstrequal{#1}{B}{ Subscripts}{%
  \ifstrequal{#1}{P}{ Notation}{%
  \ifstrequal{#1}{A}{ Acronyms}{}}}}]
}

\makenomenclature



\usepackage{dblfloatfix}
\usepackage{float}
\usepackage{cuted} 
\usepackage{caption} 

\definecolor{block-gray}{gray}{0.95}

\usepackage[framemethod=TikZ]{mdframed}

\usepackage{xpatch}

\makeatletter
\xpatchcmd{\endmdframed}
  {\aftergroup\endmdf@trivlist\color@endgroup}
  {\endmdf@trivlist\color@endgroup\@doendpe}
  {}{}
\makeatother

\usepackage{fontawesome}

\usepackage{titlesec}

\usepackage{hyperref}

\newcommand{\doi}[1]{{doi:~\href{http://doi.org/#1}{#1}}\rmFullStop}

\newcommand*{\rmFullStop}{\rmifnextchar{.}{}{}}

\makeatletter
\newcommand{\rmifnextchar}[3]{%
  \begingroup
  \ltx@LocToksA{\endgroup#2}%
  \ltx@LocToksB{\endgroup#3}%
  \ltx@ifnextchar{#1}{%
    \def\next{\the\ltx@LocToksA}%
    \afterassignment\next
    \let\scratch= %
  }{%
    \the\ltx@LocToksB
  }%
}
\makeatother

\usepackage{colortbl}

\definecolor{light-gray}{gray}{0.6}

\newcommand{\xsection}[1]{\section[#1]{\MakeUppercase{#1}}}

\usepackage{algorithm2e}
\usepackage{xcolor}

\SetAlFnt{\footnotesize}

\SetAlCapSty{xAlCapSty}

\definecolor{commentcolor}{HTML}{1E4D2B}

\SetCommentSty{xCommentSty}

\SetNlSty{mynlfont}{}{} 

\LinesNumbered

\SetSideCommentRight

\DontPrintSemicolon

\RestyleAlgo{algoruled}

\usepackage{algorithm2e}
\usepackage{etoolbox}
\usepackage{xstring}
\usepackage{calc}
 
\newlength{\xalgowidth}
\newlength{\xalgoremainder}
\newlength{\xindentwidth}

\newenvironment{vAlgorithm*}[3][]{
  \setlength{\xalgowidth}{#2} 
  \setlength{\xindentwidth}{#3} 
  \setlength{\xalgoremainder}{\textwidth-\xalgowidth} 
  \SetCustomAlgoRuledWidth{\xalgowidth} 
  \IncMargin{\xindentwidth}
  \begin{algorithm*}[#1]
}
{
  \end{algorithm*} 
  \DecMargin{\xindentwidth}
}

{
  \end{algorithm} 
  \DecMargin{\xindentwidth}
}

\makeatletter
\patchcmd{\@algocf@start}{%
\begin{lrbox}{\algocf@algobox}%
}{%
\rule{0.5\xalgoremainder}{\z@}
\begin{lrbox}{\algocf@algobox}%
\begin{minipage}{\xalgowidth}%
}{}{}
\patchcmd{\@algocf@finish}{%
\end{lrbox}%
}{%
\end{minipage}%
\end{lrbox}%
}{}{}
\makeatother

\usepackage{accents}

\definecolor{needcolor}{HTML}{C62828}

\newcolumntype{x}[1]{>{\raggedright\arraybackslash}p{#1}}

\confshortname{}
\conffullname{}
\confdate{}
\confmonth{}
\confyear{}
\confcity{}
\confcountry{}
\papernum{}

\title{Design-OS: A Specification-Driven Framework for Engineering System Design with a Control-Systems Design Case}

\author{H.~Sinan~Bank\thanks{Corresponding author: \href{mailto:sinan.bank@colostate.edu}{sinan.bank@colostate.edu}}
\affiliation{
Department of Systems Engineering \\
Colorado State University \\
Fort Collins, CO 80523 \\
}}

\author{Daniel~R.~Herber
\affiliation{
Department of Systems Engineering \\
Colorado State University \\
Fort Collins, CO 80523 \\
}}

\author{Thomas~H.~Bradley
\affiliation{
Department of Systems Engineering \\
Colorado State University \\
Fort Collins, CO 80523 \\
}}

\usepackage{amsmath}
\usepackage{tikz}
\usetikzlibrary{positioning,arrows.meta,shapes.geometric}

\usepackage{microtype}

\makeatletter
\def\@maketitle{%
  \newpage
  \null\vspace*{-26pt}%
   \vbox{\hbox to \textwidth{\begin{tabular}{@{}c@{\hskip3pc}}
   \hbox to 46pt{\vbox to 46pt{\vss\hsize46pt\vss}}\end{tabular}\hss
  \vbox{\hsize37pc\scriptsize\sf\vskip\baselineskip%
  \bannerfnt\begin{flushright}%
  \ifx\@conffullname\empty\else\hfill \@conffullname\par\fi
  \ifx\@confshortname\empty\else\hfill \@confshortname\par\fi
  \vskip.036in
  \ifx\@confmonth\empty\else\hfill \@confmonth\ \@confdate, \@confyear, \@confcity, \@confcountry\fi
  \vskip.5in
  \ifx\@papernum\empty\else\hfill {\pnumfnt\@papernum}\fi\end{flushright}}
  \hskip1pc}}
  \vskip.15in%
  \vskip .25pc{\large\twlsfb
               \begin{center}\leftskip.5in plus1fill\rightskip\leftskip
               \@title\par\end{center}}
  \vskip2pc{\begin{center}\@author\par\end{center}}\vskip12pt}
\makeatother

\begin{document}
 \setlength{\parskip}{0pt}
 \setlength{\parsep}{0pt}
 \setlength{\headsep}{0pt}
 \setlength{\topsep}{0pt}

\abovedisplayshortskip=3pt
\belowdisplayshortskip=3pt
\abovedisplayskip=3pt
\belowdisplayskip=3pt

\titlespacing*{\section}{0pt}{18pt plus 1pt minus 1pt}{3pt plus 0.5pt minus 0.5pt}

\titlespacing*{\subsection}{0pt}{9pt plus 1pt minus 0.5pt}{1pt plus 0.5pt minus 0.5pt}

\titlespacing*{\subsubsection}{0pt}{9pt plus 1pt minus 0.5pt}{1pt plus 0.5pt minus 0.5pt}

\microtypesetup{nopatch=item}
\maketitle
\microtypesetup{patch=item}


\begin{abstract}\noindent
\textit{Engineering system design---whether mechatronic, control, or embedded---often proceeds in an ad hoc manner, with requirements left implicit and traceability from intent to parameters largely absent. Existing specification-driven and systematic design methods mostly target software, and AI-assisted tools tend to enter the workflow at solution generation rather than at problem framing. Human--AI collaboration in the design of physical systems remains underexplored. This paper presents Design-OS, a lightweight, specification-driven workflow for engineering system design organized in five stages: concept definition, literature survey, conceptual design, requirements definition, and design definition. Specifications serve as the shared contract between human designers and AI agents; each stage produces structured artifacts that maintain traceability and support agent-augmented execution. We position Design-OS relative to requirements-driven design, systematic design frameworks, and AI-assisted design pipelines, and demonstrate it on a control systems design case using two rotary inverted pendulum platforms---an open-source SimpleFOC reaction wheel and a commercial Quanser Furuta pendulum---showing how the same specification-driven workflow accommodates fundamentally different implementations. A blank template and the full design-case artifacts are shared in a public repository to support reproducibility and reuse. The workflow makes the design process visible and auditable, and extends specification-driven orchestration of AI from software to physical engineering system design.}
\end{abstract}

\vspace{1ex}

\xsection{Introduction}\label{sec:introduction}

When design is ad hoc or implementation-first, requirements tend to remain implicit, and traceability from intent to parameters is difficult to achieve. AI-assisted tools compound this tendency: evidence from the design literature indicates they are predominantly applied in later solution-generation stages rather than in problem framing and specification \cite{Lee2024}. Yet the same AI capabilities, when guided by structured specifications, can support better and earlier alignment on goals---recent work on conceptual systems engineering illustrates the value of human-driven baselines alongside agentic frameworks \cite{Massoudi2025}. What is missing is an orchestration layer for the design process itself: a lightweight, model-agnostic structure that sequences design activities, enforces specification-before-implementation discipline, and provides a common interface through which both human designers and AI agents operate. This paper presents \emph{Design-OS}, a lightweight, specification-driven, phased workflow that addresses this gap by making specifications the binding contract between conceptual intent and implementation and by front-loading literature and conceptual design before committing to a specification. \emph{Design-OS} adopts this collaborative, specification-first stance while remaining compatible with agentic tools where they add value, and structures the process so that traceability from intent to parameters can be agnostically documented and converted for other schemas or frameworks.

\emph{Design-OS} proceeds in five stages: concept definition $\to$ literature survey $\to$ conceptual design $\to$ requirements definition $\to$ design definition, with selected stage terminology aligned with INCOSE SE Handbook processes (notably Design Definition and Requirements Definition)~\cite{INCOSE2023}. It is intended for engineering design, research prototypes, education, and AI-augmented design workflows. By using a control system design project as a design case, we demonstrate how the workflow supports the kind of structured design process and traceability that engineering design methodology calls for \cite{Pahl2007}, in a setting where prior work on this benchmark has focused largely on control algorithms and performance analysis \cite{Hamza2019} rather than on embedding the design process in a formal, traceable workflow. This paper addresses three research questions:
\begin{list}{}{\setlength{\leftmargin}{0pt}\setlength{\itemindent}{0pt}\setlength{\listparindent}{0pt}\setlength{\itemsep}{2pt}\setlength{\parsep}{0pt}\setlength{\topsep}{4pt}}
\item \textbf{RQ1.} Why is a lightweight, specification-driven workflow needed for physical engineering system design? (Sec.~\ref{sec:introduction}--\ref{sec:relatedwork})
\item \textbf{RQ2.} How does \emph{Design-OS} relate to existing specification-driven, systematic-design, and AI-assisted design frameworks? (Sec.~\ref{sec:relatedwork}--\ref{sec:methodology})
\item \textbf{RQ3.} How does the workflow support traceability from intent to parameters across domains? (Sec.~\ref{sec:usecase}--\ref{sec:discussion})
\end{list}

Aligning with these questions, the contributions of this paper are: (1)~a concise presentation of the \emph{Design-OS} methodology and its positioning relative to specification-driven and requirements-driven design (including industry practice such as GitHub Spec Kit~\cite{GitHubSpecKit} and Microsoft specification-driven development~\cite{MicrosoftSDD}), systematic design frameworks (Pahl \& Beitz~\cite{Pahl2007}, VDI~\cite{Graessler2020}, Stage-Gate~\cite{Cooper1990}), and AI-assisted design. (2)~An end-to-end design case showing how the workflow was run and how artifacts trace from context and literature through to requirements and implementation plan. (3)~A discussion of implications for design education and for specification-driven orchestration of AI in physical engineering system design.

The remainder of the paper is organized as follows: Section~\ref{sec:relatedwork} reviews related work. Section~\ref{sec:methodology} describes the \emph{Design-OS} methodology. Section~\ref{sec:usecase} presents the inverted pendulum design case. Section~\ref{sec:discussion} discusses limitations and future work, and Section~\ref{sec:conclusion} concludes.

\begin{figure*}[t]
\centering
\includegraphics[width=1.\textwidth]{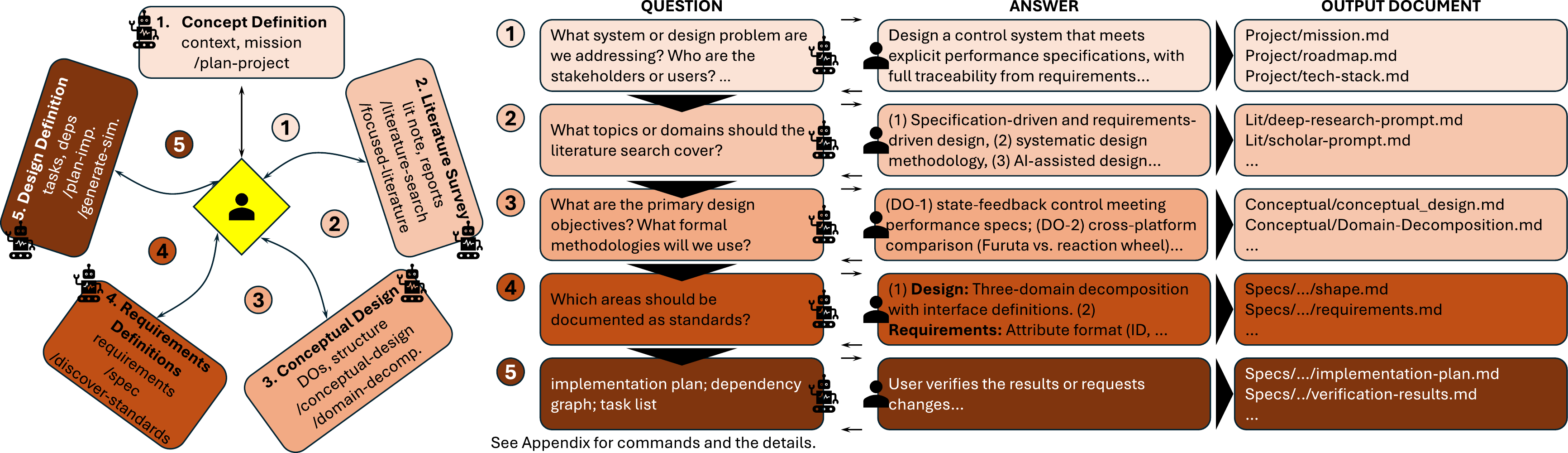}
\caption{\emph{Design-OS} workflow (abstract): five stages arranged around the designer (central diamond), each executed by an AI agent (robot icon). User interactions (numbered) between stages enable validation and iteration. Each box shows the stage name, key artifacts, and commands.}
\label{fig:design-os-stages}
\end{figure*}


\xsection{Related Work}\label{sec:relatedwork}

\paragraph{Specification-driven and requirements-driven design.}
The principle that formal specifications must precede implementation is well established in requirements engineering and systems engineering \cite{Nuseibeh2000,Zave1997}. Traceability from requirements to design and verification is central; inadequacy of pre-requirements specification traceability is a major source of failure \cite{Gotel1994}. Reference models for traceability have since formalized the relationships among requirements, design decisions, and verification artifacts \cite{Ramesh2001}. Standards such as IEEE~830 \cite{IEEE830} and its successor ISO/IEC/IEEE~29148 \cite{ISO29148} define quality criteria and templates for software and system requirements specifications. Structured syntax approaches such as EARS provide five requirement templates (precondition--trigger--shall--response) to reduce ambiguity and vagueness in natural-language requirements \cite{Mavin2009}. Industry practice has recently emphasized specification-driven development (SDD) in the age of AI coding assistants, treating specifications as the source of truth from which code is generated or verified rather than the reverse \cite{Piskala2026}: \textbf{Microsoft} promotes specifications as ``version control for your thinking,'' with technical decisions made explicit and reviewable before code becomes the de facto specification \cite{MicrosoftSDD}. The open-source \textbf{GitHub Spec Kit} provides a toolkit for outlining requirements, motivations, and technical aspects before handing work to AI agents (e.g., GitHub Copilot), including a ``Constitution'' for engineering guidelines, clear specifications, technical plans, and implementation tasks \cite{GitHubSpecKit}. \emph{Design-OS} aligns with this philosophy and extends it from software to \emph{physical} engineering system design, with stages---literature review, conceptual design of mechatronic or control systems, performance specifications---that are domain-specific.

\paragraph{Systematic design methodology.}
\emph{Design-OS}'s phased workflow draws on systematic design frameworks. Pahl and Beitz's four-phase model (task clarification, conceptual design, embodiment design, detail design) is reflected in \emph{Design-OS}'s early phases \cite{Pahl2007}. Gero's Function-Behavior-Structure (FBS) knowledge representation schema provides a complementary theoretical lens, formalizing how design transforms functions into structural descriptions through expected and actual behaviors \cite{Gero1990}. VDI~2206 (V-model for mechatronics and cyber-physical systems) prescribes phased deliverables and continuous requirements management \cite{Graessler2020}. \emph{Design-OS} aims to compress these into a workflow optimized for specification-driven engineering. Stage-Gate \cite{Cooper1990} shares the idea of discrete phases with explicit deliverables and front-loading. \emph{Design-OS} focuses on the technical design workflow rather than the full product lifecycle. \emph{Design-OS} does not replace these frameworks but offers a lightweight, AI-friendly instantiation that can be used on its own or alongside MBSE and VDI, where a lighter-weight, specification-driven workflow is desired.

\paragraph{AI-assisted design and the specification-driven gap.}
Large language models and multi-agent frameworks are reshaping design workflows. Multi-agent systems such as MetaGPT and ChatDev use structured intermediate outputs (requirement documents, system designs, interface specs) to orchestrate agents and reduce cascading errors \cite{Hong2024,Qian2024}---directly analogous to \emph{Design-OS}'s phased artifacts. A notable gap is that specification-driven orchestration of AI agents for \emph{physical} engineering system design has received little attention. Existing multi-agent pipelines target software development \cite{Zadenoori2025}. Recent work has shown that LLMs can generate conceptual design solutions with higher average feasibility and usefulness than crowdsourced alternatives, though crowdsourced solutions exhibit more novelty \cite{Ma2023}, and that generative pre-trained transformers can produce diverse design concepts from textual prompts \cite{Zhu2023}---yet these capabilities remain disconnected from structured specification workflows. Studies of human--AI collaboration in design indicate that most AI design tools support later solution-generating stages, with few supporting early-stage problem discovery \cite{Lee2024}. \emph{Design-OS} addresses this gap by providing a workflow that spans from concept definition and literature survey through to design definition, extending the specification-driven multi-agent paradigm to physical engineering design.

\paragraph{Control education and the design case.}
Control systems engineering and education have long relied on canonical benchmark platforms---the inverted pendulum, ball-and-beam, ball-and-plate, and servo systems---to teach dynamics, underactuation, and feedback design \cite{Awtar2002,Boubaker2013}. Commercial platforms are widely deployed, and open-source alternatives are also available. Control education methodology emphasizes interactivity and structured use of technology \cite{Dormido2004}. Yet these systems are rarely situated within a \emph{formal, traceable} design process: the mapping from performance specifications to controller parameters is typically an in-paper exercise rather than a documented workflow.
\emph{Design-OS} targets this gap by providing a specification-driven workflow applicable to any of these platforms. We demonstrate it on two inverted pendulum platforms (Furuta and reaction wheel) and share a blank template that can be adapted to other projects.

\xsection{Methodology: Design-OS}\label{sec:methodology}

\emph{Design-OS} is a specification-driven, phased workflow for engineering system design. Each stage produces explicit artifacts that inform the next, supporting a path to traceability from initial motivations to the final implementation plan (realizing such traceability in practice remains challenging and tool-dependent). The workflow is designed to be lightweight enough for research projects, capstone design, and rapid iteration, while retaining the rigor of formal requirements management.

\paragraph{Stage 1:~Concept Definition.}
The designer captures context, mission, and scope in a structured project folder (e.g., \texttt{mission.md}, \texttt{roadmap.md}, \texttt{tech-stack.md}). This stage establishes the problem statement, stakeholders, success criteria, and constraints. It parallels the Business or Mission Analysis process in INCOSE~\cite{INCOSE2023}, task clarification in Pahl \& Beitz~\cite{Pahl2007}, and problem identification in DSRM~\cite{Peffers2007}.

\paragraph{Stage 2:~Literature Survey.}
An initial and focused literature search is conducted, aligned with the project goals. Artifacts include a preliminary literature note, deep-research prompts, scholar search queries, and literature reports. The designer formulates structured research prompts describing the topics and gaps to investigate. AI research assistants (e.g., ChatGPT Deep Research, Claude, Gemini) then execute these prompts and \emph{return} candidate references, summaries, and gap analyses. The designer curates the results---verifying that cited works exist, removing hallucinated references, and merging findings into the preliminary literature note. The key outcome is a landscape of prior work, identified gaps, and references that feed conceptual design. This stage enforces that design decisions are grounded in existing knowledge before any specification is written.

\paragraph{Stage 3:~Conceptual Design.}
Design objectives (DOs), product or publication structure, and a positioning narrative are developed. The designer determines what the artifact is, how it relates to prior work, and what gap it fills. In a control design context, this stage identifies candidate modeling approaches, control strategies, and system architectures. The conceptual design is fixed before committing to the requirements definition, reducing downstream rework.

\paragraph{Stage 4:~Requirements Definition.}
Stage~4 proceeds in two substeps: \emph{shape} (define section-level requirements and structure) and \emph{elaborate} (fill in detailed requirements with IDs, ``shall'' statements, source, priority, and verification method). The specification also documents functions, architecture, interfaces, and validation criteria. The specification is the \emph{binding contract} between intent and implementation: it serves as the single source of truth that governs all downstream work. Standards discovered during the process (e.g., requirements format, system structure templates) are documented alongside the specification.

\paragraph{Stage 5:~Design Definition.}
A task breakdown, dependency ordering, and execution plan are derived from the specification. Each task maps to a specific deliverable and to the requirement(s) it satisfies. In a control design project, the plan might include: derive a mathematical model, design controller gains from requirements, build a simulation, run verification, and document traceability.

\paragraph{Traceability across stages.}
Traceability is maintained by linking requirement IDs to design parameters and to verification steps (e.g., REQ-001 $\leftrightarrow$ settling time $\leq$ 1.0\,s $\leftrightarrow$ state-feedback gain $K$ $\leftrightarrow$ simulation step response). The process is iterative within stages, but stage order is respected: no design definition is written before an agreed requirements definition, and no requirements definition before a conceptual design. Figure~\ref{fig:design-os-stages} summarizes the stages and their main artifacts.


\xsection{Design Case: Inverted Pendulum}\label{sec:usecase}

We applied the \emph{Design-OS} workflow to an inverted pendulum control design project---a canonical benchmark in control education and research \cite{Furuta1992,Hamza2019}.

The AI-assisted literature search (Stage~2) mapped a landscape of commercial and open-source platforms for control education, including Quanser (rotary and linear pendulums) \cite{QuanserResearchPapers}, Feedback Instruments (Digital Pendulum) \cite{FeedbackDigitalPendulum}, and several open-hardware projects (SimpleFOC reaction wheel pendulum, 3D-printable Furuta pendulums, cart-pole kits). From this landscape, two platforms were selected at the feasibility gate (Stage~3): (1)~the \textbf{Quanser SRV02} Furuta-type rotary inverted pendulum, chosen for its extensive academic track record---deployed in hundreds of university labs worldwide with ABET-aligned courseware and manufacturer-documented parameters---and (2)~the \textbf{SimpleFOC-based reaction wheel inverted pendulum} \cite{SimpleFOCPendulum,Skuric2022}, chosen for its open-source hardware and software, comprehensive build documentation, active community, and validated control examples \cite{Wright2023}. This selection demonstrates that the same specification-driven workflow and requirements can be satisfied by fundamentally different implementations---one commercial, one open-source. Full artifacts for this design case, along with a blank \emph{Design-OS} template for other potential projects, are available in a public GitHub repository~\cite{DesignOS2026}.

\subsection{Stage 1--2: Concept Definition and Literature Survey}

\paragraph{Context.}
The mission was defined as follows: design a control system for a rotary inverted pendulum that meets explicit performance specifications, with full traceability from requirements to controller parameters across the mechanical, electrical, and software domains. The roadmap established five phases (concept definition, literature survey, conceptual design, requirements definition, design definition), and the tech stack was defined as MATLAB/Simulink for modeling and control, with Python as an alternative.

\paragraph{Literature.}
A literature search targeted four areas: (i)~specification-driven and requirements-driven design, (ii)~systematic design methodology, (iii)~AI-assisted design, and (iv)~control education and the inverted pendulum benchmark. Structured deep-research prompts were formulated and executed by multiple AI research assistants (Claude, Gemini). Each assistant independently searched for and returned candidate references, which were then verified by the designer and merged into a preliminary literature note. The designer specified the research topics; the AI assistants independently identified the specific references within those topics. The following key papers were \emph{surfaced by the AI-assisted search} and, after verification, informed the conceptual design: Furuta et al.~\cite{Furuta1992} for the original Furuta pendulum, Cazzolato and Prime \cite{Cazzolato2011} for the definitive dynamics derivation, Hamza et al.~\cite{Hamza2019} for the benchmark survey, Quanser's rotary pendulum workbook for hardware parameters and lab specifications, Skuric et al.~\cite{Skuric2022} for the SimpleFOC library and its example implementation underlying the open-source reaction wheel platform \cite{SimpleFOCPendulum}, and Wright \cite{Wright2023} for a validated reaction wheel pendulum platform designed for controls education using SimpleFOC.

\subsection{Stage 3: Conceptual Design and Domain Decomposition}

Three design objectives were defined:
\begin{list}{}{%
\setlength{\leftmargin}{0pt}\setlength{\itemindent}{0pt}\setlength{\listparindent}{0pt}\setlength{\itemsep}{2pt}\setlength{\parsep}{0pt}\setlength{\topsep}{4pt}}

\item \textbf{(DO-1)}~What state-feedback control design satisfies the performance specifications for each platform? 
    
\item \textbf{(DO-2)}~How do fundamentally different platforms (SimpleFOC reaction wheel vs.\ Quanser Furuta) satisfy the same set of requirements?

\item \textbf{(DO-3)}~How does cross-domain traceability flow from performance specifications through software, electrical, and mechanical domains to controller parameters?

\end{list}

\begin{figure}[t]
\centering
\includegraphics[width=\columnwidth]{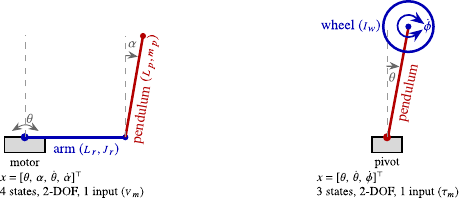}
\caption{Two inverted pendulum architectures: Furuta (left) and reaction wheel (right).}
\label{fig:pendulum-architectures}
\end{figure}

 Two system architectures were identified (Figure~\ref{fig:pendulum-architectures}): (a)~a Furuta-type rotary inverted pendulum with state vector $x = [\theta,\; \alpha,\; \dot{\theta},\; \dot{\alpha}]^\top$ (4~states, 2-DOF), and (b)~a reaction wheel inverted pendulum with state vector $x = [\theta,\; \dot{\theta},\; \dot{\phi}]^\top$ (3~states, 2-DOF), where $\theta$ is the pendulum angle and $\dot{\phi}$ is the reaction wheel angular velocity \cite{Wright2023}.

\paragraph{Literature-driven domain discovery.}
The domain decomposition was driven by the literature \emph{surfaced} in Stage~2---the designer specified the topics, and the AI assistants returned the specific references that established each domain. Furuta et al.\ \cite{Furuta1992} and Cazzolato and Prime \cite{Cazzolato2011} describe the two-link arm--pendulum system with Euler--Lagrange formulation, establishing the \emph{mechanical} domain. The Quanser SRV02 documentation describes the DC servo motor, gearbox, encoders, and power amplifier---establishing the \emph{electrical} domain and the torque equation in Eq.~\eqref{eq:torque} that couples it to mechanical torque. The control design literature (pole placement, state feedback) and simulation environment (MATLAB/Simulink) establish the \emph{software} domain. Where literature was ambiguous (e.g., whether the DAQ interface board belongs to electrical or software), the designer was consulted at a user checkpoint. The three domains identified are:
\begin{list}{}{%
\setlength{\leftmargin}{0pt}\setlength{\itemindent}{0pt}\setlength{\listparindent}{0pt}\setlength{\itemsep}{2pt}\setlength{\parsep}{0pt}\setlength{\topsep}{4pt}}

\item \textbf{Mechanical:} rotary arm (length $L_r$, inertia $J_r$, damping $B_r$) and pendulum link (length $L_p$, mass $m_p$, inertia $J_p$, damping $B_p$). Two-link dynamics derived via the Euler--Lagrange equations.

\item \textbf{Electrical:} DC servo motor ($R_m$, $k_m$, $k_t$), gearbox ($K_g$, $\eta_g$, $\eta_m$), power amplifier, and rotary encoders for $\theta$ and $\alpha$. The torque equation:
\begin{equation}\label{eq:torque}
\tau = \frac{\eta_g K_g \eta_m k_t (V_m - K_g k_m \dot{\theta})}{R_m}
\end{equation}
couples the electrical domain (voltage $V_m$) to the mechanical domain (torque $\tau$).

\item \textbf{Software:} state-feedback control law $u = K(x_d - x)$, state estimation from encoders via differentiation with high-pass filtering, controller enable/disable logic, simulation environment.

\end{list}

The state-space matrices $A$ and $B$ encode the coupling: rows~3--4 of $A$ and $B$ incorporate the motor torque equation in Eq.~\eqref{eq:torque}, making the electrical parameters ($R_m$, $k_m$, $k_t$, $K_g$) part of the plant model. The same three-domain decomposition applies to the reaction wheel pendulum, with different specifics: the mechanical domain is a single-link pendulum with a reaction wheel (inertia $I_w$) instead of a rotary arm. The electrical domain uses a BLDC motor with field-oriented control (torque constant $K_t$, phase resistance $R$) instead of a geared DC servo. The software domain uses Arduino/C++ with the SimpleFOC library instead of MATLAB/Simulink.

\paragraph{Feasibility gate.}
Before proceeding to Requirements Definition (Stage~4), a validation gate was run. The cost comparison (Table~\ref{tab:platform}) was produced at this gate: the SimpleFOC-based reaction wheel pendulum \cite{SimpleFOCPendulum} (\$100--200, open-source Arduino/C++ stack) was compared against the Quanser SRV02 platform (\$5k+, proprietary MATLAB/QUARC licensing). The SimpleFOC platform offers accessibility while eliminating licensing constraints. The Quanser platform provides manufacturer-documented parameters and an established lab infrastructure. Wright \cite{Wright2023} demonstrated that the SimpleFOC-class platform can be fully characterized through system identification (pendulum inertia, damping, motor torque constant, phase resistance), providing the parameter certainty needed for model-based control design. Software licensing was assessed: Python and Arduino/C++ are freely available; MATLAB/Simulink is proprietary. The designer confirmed at a user checkpoint that both platforms would be carried forward: the SimpleFOC reaction wheel pendulum as the primary open-source platform, and the Quanser as a well-documented commercial comparison. A Python simulation was specified for reproducibility.

\subsection{Stage 4: Requirements Definition}

The Requirements Definition stage produced requirements organized by domain (Table~\ref{tab:requirements}). Performance requirements (REQ-01--04) are system-level. Domain-specific requirements (mechanical, electrical, software) ensure that the full mechatronic design is captured and traceable.

\begin{table}[t]
\def\arraystretch{1.2}
\centering
\caption{Requirements by domain for the rotary inverted pendulum.}
\label{tab:requirements}
\small
\begin{tabular}{@{}llx{3.4cm}l@{}}
\hline \hline
\textbf{ID} & \textbf{Domain} & \textbf{Requirement (shall)} & \textbf{Verif.} \\
\hline
REQ-01 & Perf.  & Damping ratio $\zeta = 0.7$ & Sim. \\
REQ-02 & Perf.  & Natural freq.\ $\omega_n = 4$\,rad/s & Sim. \\
REQ-03 & Perf.  & $|\alpha| < 15$\,deg & Sim. \\
REQ-04 & Perf.  & $|V_m| < 10$\,V & Sim. \\
REQ-05 & Mech.  & State-space from first-principles EOM + linearization & Review \\
REQ-06 & Mech.  & Parameters documented with source & Review \\
REQ-07 & Elec.  & Actuator dynamics ($R_m$, $k_m$, $k_t$, $K_g$) in model & Review \\
REQ-08 & Elec.  & Encoders measure $\theta$, $\alpha$ & Review \\
REQ-09 & SW     & Control at sufficient sampling rate & Review \\
REQ-10 & SW     & Velocity estimation with filtering & Review \\
REQ-11 & SW     & Gains from requirements (pole placement) & Review \\
REQ-12 & All    & Traceability table: REQ $\to$ param $\to$ verif. & Review \\
\hline \hline
\end{tabular}
\end{table}

\paragraph{Modeling (mechanical + electrical).}
The nonlinear equations of motion are derived via the Euler--Lagrange method. Letting $q = [\theta,\, \alpha]^\top$, the dynamics take the matrix form:
\begin{equation}\label{eq:eom-matrix}
  D(q)\,\ddot{q} + C(q,\dot{q})\,\dot{q} + g(q) = \tau,
\end{equation}
where $D$ is the inertial matrix, $C$ the damping/Coriolis matrix, $g$ the gravitational vector, and $\tau$ the torque vector. After linearization about the upright equilibrium and incorporation of the motor torque in Eq.~\eqref{eq:torque} (REQ-05, REQ-07), the linear state-space model is:
\begin{equation}\label{eq:ss}
  \dot{x} = Ax + Bu, \qquad y = Cx + Du,
\end{equation}
where $u = V_m$. Using Quanser SRV02 parameters:
\begin{equation}\label{eq:Amat}
A = \begin{bmatrix}
0 & 0 & 1 & 0 \\
0 & 0 & 0 & 1 \\
0 & 80.3 & -45.8 & -0.93 \\
0 & 122 & -44.1 & -1.40
\end{bmatrix}\!,\quad
B = \begin{bmatrix}
0 \\ 0 \\ 83.4 \\ 80.3
\end{bmatrix}\!.
\end{equation}

The entries in rows~3--4 reflect the coupling of mechanical parameters ($m_p$, $L_p$, $J_r$, $J_p$) with electrical parameters ($R_m$, $k_m$, $k_t$, $K_g$). The open-loop poles $\{7.12,\; 0,\; -5.95,\; -48.37\}$ confirm instability.

\paragraph{Reaction wheel pendulum model.}
For the SimpleFOC reaction wheel pendulum, the state vector is $x = [\theta,\; \dot{\theta},\; \dot{\phi}]^\top$ with input $u = \tau_m$ (motor torque). Wright \cite{Wright2023} derived the linearized model via Lagrangian mechanics and identified parameters through system identification (oscillation period $\to$ inertia, logarithmic decrement $\to$ damping, torque--current characterization $\to$ $K_t$). The resulting state-space matrices are:
\begin{equation}\label{eq:Amat-rw}
A_{\mathrm{rw}} = \begin{bmatrix}
0 & 1 & 0 \\
30.86 & -0.58 & 0 \\
-30.86 & 0.58 & 0
\end{bmatrix}\!,\quad
B_{\mathrm{rw}} = \begin{bmatrix}
0 \\ -127.9 \\ 5652.8
\end{bmatrix}\!.
\end{equation}
The structure mirrors the Furuta model: the $A$ matrix encodes the gravitational instability (positive entry $a_{21} = 30.86$) and the $B$ matrix encodes the motor torque coupling, but with three states instead of four. This illustrates how the same \emph{Design-OS} workflow produces analogous but distinct state-space models for different platforms---both traceable to their respective parameters and requirements.

\paragraph{Control design (software domain, REQ-11).}
A state-feedback controller $u = K(x_d - x)$ is designed via pole placement. The dominant closed-loop poles are placed at $p_{1,2} = -\sigma \pm j\omega_d$ where $\sigma = \zeta\omega_n = 2.8$ and $\omega_d = \omega_n\sqrt{1-\zeta^2} = 2.86$\,rad/s, satisfying REQ-01 and REQ-02. The gain vector $K$ is computed analytically from the desired characteristic polynomial, ensuring traceability from performance specifications through the gain computation to the closed-loop response (REQ-11).

\paragraph{Simulation verification (REQ-01--04).}
A Python simulation was generated by the \emph{Design-OS} \texttt{/generate-simulation} command directly from the specification artifacts. The script builds the state-space model from documented parameters, computes gain~$K$ via pole placement, and simulates the closed-loop step response. Table~\ref{tab:verification} summarizes the verification results: all four performance requirements pass. The simulation code, plots, and verification report are included in the repository alongside the specification artifacts, completing the traceability chain from requirements through design parameters to numerical verification.

\begin{table}[t]
\def\arraystretch{1.2}
\centering
\caption{Simulation verification of performance requirements.}
\label{tab:verification}
\small
\begin{tabular}{@{}llll@{}}
\hline \hline
\textbf{REQ} & \textbf{Specification} & \textbf{Simulation} & \textbf{Result} \\
\hline
01 & $\zeta = 0.7$ & $\zeta = 0.700$ & Pass \\
02 & $\omega_n = 4$\,rad/s & $\omega_n = 4.000$\,rad/s & Pass \\
03 & $|\alpha| < 15^\circ$ & $|\alpha|_{\max} = 2.9^\circ$ & Pass \\
04 & $|V_m| < 10$\,V & $|V_m|_{\max} = 0.5$\,V & Pass \\
\hline \hline
\end{tabular}
\end{table}

\subsection{Cross-Domain Traceability and Platform Comparison}

\paragraph{Cross-domain requirement flow.}
A key insight is that performance requirements flow through all three domains. For example, REQ-01 ($\zeta = 0.7$) determines the closed-loop pole locations (software: pole placement $\to$ gain $K$), which determine the voltage command $V_m$ (electrical: motor equation), which produces torque $\tau$ (mechanical: arm and pendulum dynamics). Table~\ref{tab:traceability} summarizes this cross-domain traceability.

\begin{table}[t]
\def\arraystretch{1.2}
\centering
\caption{Cross-domain traceability: requirement $\to$ domain flow $\to$ verification.}
\label{tab:traceability}
\small
\begin{tabular}{lx{4.2cm}x{2.3cm}}
\hline \hline
\textbf{REQ} & \textbf{Domain flow} & \textbf{Verification} \\
\hline
01, 02 & Perf $\to$ SW (gain $K$) $\to$ Elec ($V_m$) $\to$ Mech ($\tau$) & Sim.\ step resp. \\
03     & Perf $\to$ Mech (pendulum $\alpha$) & Sim.\ deflection \\
04     & Perf $\to$ Elec (amplifier limit) & Sim.\ $V_m$ \\
05     & Mech + Elec $\to$ $A, B$ matrices & Derivation \\
07     & Elec ($R_m, k_m, k_t, K_g$) $\to$ $A, B$ & Review \\
11     & SW (pole placement) $\to$ $K$ & Design rationale \\
\hline \hline
\end{tabular}
\end{table}

\paragraph{Platform comparison (feasibility gate output).}
As produced by the feasibility gate in Stage~3, Table~\ref{tab:platform} compares two implementations satisfying the same requirements: (1)~the SimpleFOC-based reaction wheel pendulum (open-source, BLDC motor, Arduino) \cite{SimpleFOCPendulum}, and (2)~the Quanser SRV02 with Furuta pendulum (commercial, proprietary, DC servo). Both satisfy the performance requirements (REQ-01--04), but through different mechanical configurations (reaction wheel vs.\ arm-driven Furuta, both 2-DOF), different electrical subsystems (BLDC with field-oriented control vs.\ DC servo), and different software stacks (Arduino/C++ vs.\ MATLAB/Simulink). The domain-specific requirements (REQ-05--12) make these differences explicit and traceable. Wright \cite{Wright2023} validated a SimpleFOC-class reaction wheel pendulum with full system identification and state-feedback control, demonstrating that the open-source platform can achieve model-based control design comparable to commercial platforms. The feasibility gate found that the SimpleFOC platform costs $\sim$50$\times$ less and removes licensing constraints, while requiring system identification for parameter characterization.

\begin{table}[t]
\def\arraystretch{1.2}
\centering
\caption{Platform comparison: same requirements, different implementations.}
\label{tab:platform}
\small
\begin{tabular}{lx{3.0cm}x{3.4cm}}
\hline \hline
\textbf{Domain} & \textbf{SimpleFOC} & \textbf{Quanser SRV02} \\
\hline
Mechanical & 3D-printed, sys.\ ID needed \cite{Wright2023} & Aluminum arm + pendulum, known params \\
Electrical & BLDC + FOC driver, Arduino & DC servo, gearbox, commercial DAQ \\
Software   & C++ + SimpleFOC lib. & MATLAB/Simulink, QUARC \\
Pendulum   & Reaction wheel (2-DOF) & Furuta (2-DOF) \\
Cost       & $\sim$\$100--200 & $\sim$\$5k+ \\
\hline \hline
\end{tabular}
\end{table}

\subsection{AI Model Comparison}

To assess whether the \emph{Design-OS} workflow produces consistent results across AI models, we ran the same pipeline (from plan-project through generate-simulation) with two proprietary models: Claude Opus~4.6 (Anthropic) and Gemini~3.1~Pro (Google). This comparison evaluates which stages each model handles well, where outputs diverge, and whether the structured commands are sufficient to guide consistent results across models. Table~\ref{tab:model-stage} rates each model's output quality per \emph{Design-OS} stage; Table~\ref{tab:model-corrections} summarizes the type and frequency of human corrections required.

Early indicators suggest that the structured command format and explicit user checkpoints reduce model-dependent variation. 


\begin{table*}[t]
\def\arraystretch{1.2}
\centering
\caption{AI model output quality by \emph{Design-OS} stage. Source data: \texttt{log/report\_claude.md}, \texttt{log/report\_gemini.md}.}
\label{tab:model-stage}
\small
\begin{tabular}{x{3.4cm}x{6.7cm}x{6.7cm}}
\hline \hline
\textbf{} & \textbf{Claude Opus 4.6} & \textbf{Gemini 3.1 Pro} \\
\hline
\textbf{Overview} &
  36 artifacts, 29 checkpoints, 24 corrections. Python (model choice). Platforms: Quanser SRV02 + SimpleFOC reaction wheel. &
  $\sim$25 artifacts, 28 checkpoints, 14 corrections. MATLAB (tech-stack). Platforms: Quanser QUBE + ESP32. \\
\hline
\textbf{Stage 1: Concept Def.} &
  Complete --- 0 errors. Mission, roadmap, tech stack correct; open tech stack appropriately deferred. &
  Complete --- 0 errors. Mission, roadmap, tech stack correct; MATLAB explicitly committed. \\
\hline
\textbf{Stage 2: Lit. Survey} &
  Good --- 21 ref.\ errors caught by \texttt{/verify-references}; 8-theme coverage; pipeline step self-initiated. &
  Fair --- 55 sources; 2 citation mismatches found by post-hoc \texttt{/verify-references} (2026-03-07); step not self-initiated. \\
\hline
\textbf{Stage 3: Concept. Design} &
  Good --- 1 DOF factual error (1-DOF stated, 2-DOF correct); platform selection and functions correct. &
  Good --- 1 domain error (Control/Math domain missing from decomposition); platforms, functions correct. \\
\hline
\textbf{Stage 4: Req. Def.} &
  Good --- 12 requirements, 4 formal standards (\texttt{/discover-standards}); 2 minor corrections. &
  Poor --- 7 requirements; \texttt{/validate-feasibility} gate skipped; \texttt{/discover-standards} not run; 3 specification errors: scope deviation, phantom REQ-HWW IDs, HIL in REQ-IMP-002. \\
\hline
\textbf{Stage 5: Design Def.} &
  Complete --- both platforms ALL PASS; specification-driven order maintained; 1 numerical error (open-loop poles). &
  Fair --- both platforms pass; execution order corrected; tool mismatch between MATLAB artifacts and Python report. \\
\hline \hline
\end{tabular}
\end{table*}

\begin{table}[t]
\def\arraystretch{1.2}
\centering
\caption{Human corrections needed per model.}
\label{tab:model-corrections}
\small
\begin{tabular}{@{}x{2.0cm}cc@{}}
\hline \hline
\textbf{Correction} & \textbf{Claude Opus 4.6} & \textbf{Gemini 3.1 Pro} \\
\hline
Factual errors       & 20 & 1   \\
Missing reqs.        & 0  & 1   \\
Halluc.\ refs.       & 3  & 2   \\
Struct.\ deviat.     & 1  & 9   \\
Domain errors        & 0  & 1   \\
\hline
\textit{Total}                & 24 & 14  \\
\hline \hline
\end{tabular}
\end{table}

\xsection{Discussion}\label{sec:discussion}

\paragraph{What worked.}
The \emph{Design-OS} workflow proved effective for structuring the progression from context and literature through to requirements and implementation plan. Several aspects stood out. First, the specification served as the \emph{single source of truth} and reduced ad hoc decisions during implementation: once requirements were formalized with IDs, ``shall'' statements, and verification methods, the controller design became a matter of satisfying those requirements rather than iterating by trial and error. Second, the explicit literature stage prevented premature commitment to a solution. The conceptual design was informed by a landscape of prior work rather than by the first approach that came to mind. Third, the phased structure made the design process \emph{auditable}: a reviewer (or instructor) can inspect the chain from mission to requirements to parameters to verification. Fourth, the literature-driven domain decomposition prevented the AI from fabricating system structure: each domain boundary was grounded in published work. Fifth, the validation gates surfaced cost and feasibility constraints early---particularly the Quanser vs.\ SimpleFOC comparison---preventing commitment to infeasible approaches. Sixth, user checkpoints ensured that the designer retained authority over decisions where literature was insufficient, supporting genuine human--AI teaming.

\paragraph{Comparison with typical practice.}
In typical control design coursework and many research projects, the path from performance specifications to controller gains is an in-paper or in-lab exercise: students receive specifications (e.g., damping ratio, settling time), compute gains, and report results. The mapping from intent to parameters exists but is rarely documented as a traceable workflow. \emph{Design-OS} adds a thin but explicit layer of structure---concept definition, literature survey, conceptual design, requirements definition, design definition---that makes this mapping inspectable and reproducible. Compared to full MBSE or VDI 2206 processes, \emph{Design-OS} is lighter weight and does not require specialized modeling languages (e.g., SysML, Modelica) or dedicated MBSE platforms---it operates on structured Markdown artifacts with AI models (e.g., Claude Opus~4.6) preferably inside an IDE (e.g., VS~Code). Compared to ad hoc approaches, it provides structure to support traceability and reproducibility.

\paragraph{Limitations.}
Several limitations should be noted. First, maintaining the specification introduces overhead: writing and updating requirements, traceability tables, and implementation plans takes time that would otherwise be spent on implementation. For very small projects, this overhead may not be justified. Second, the workflow requires discipline in respecting stage order; in practice, designers may be tempted to skip ahead to implementation before finalizing the specification. Third, end-to-end traceability from intent to parameters remains challenging and tool-dependent ---the workflow structures for it conceptually, but the degree to which it can be made fully explicit depends on the tools and environment. Fourth, the design case presented here is a well-understood benchmark. Applying \emph{Design-OS} to novel or poorly understood systems where requirements are harder to specify upfront remains to be validated. Fifth, the validation gates as currently implemented are lightweight checklists rather than formal analysis; for safety-critical systems, more rigorous gate criteria would be needed. Sixth, the user checkpoints rely on the designer being available and engaged; in a fully asynchronous workflow, the interruptions may disrupt flow.

\paragraph{Implications for education.}
For engineering design education, the workflow makes the design process \emph{visible} and provides a structure within which students can be assessed not only on their final controller performance but on the quality of their requirements, the traceability of their design decisions, and the completeness of their documentation. The phased structure also provides natural checkpoints for instructor feedback. Whether this structure has the potential to ease learning is not evident from this paper and would require evaluation in a classroom or controlled setting. We plan such studies as future work. The blank template shared alongside this paper can be adapted to mechatronics and control systems courses; the inverted pendulum design case includes both a low-cost SimpleFOC reaction wheel platform and a commercial Quanser Furuta platform, giving instructors flexibility in hardware selection.

\paragraph{Implications for AI-assisted design.}
\emph{Design-OS}'s phased artifacts are well-suited to AI-assisted workflows. Each stage produces structured documents (mission, literature notes, requirements, implementation plans) that can be generated or reviewed by large language models. The specification serves as a shared contract between human designers and AI agents, aligning with the philosophy of GitHub Spec Kit and Microsoft specification-driven development. Multi-agent frameworks such as MetaGPT \cite{Hong2024} could be adapted to execute \emph{Design-OS} stages, with agents assigned to literature search, conceptual design, specification elaboration, and implementation planning.

\paragraph{Future work.}
Future work includes applying \emph{Design-OS} to additional design cases beyond the inverted pendulum family, extending the AI model comparison to open-weight models (e.g., Kimi~K2.5), tightening integration with AI agents (e.g., automated literature summarization, specification-to-task decomposition, and requirement verification), and exploring tooling that better supports intent-to-parameter traceability. We also plan to evaluate the workflow in a classroom setting with student teams.
\xsection{Conclusion}\label{sec:conclusion}

We presented \emph{Design-OS}, a lightweight, specification-driven, phased workflow for engineering system design (concept definition $\to$ literature survey $\to$ conceptual design $\to$ requirements definition $\to$ design definition), enhanced with literature-driven domain decomposition, validation gates between stages, and user checkpoints for human--AI teaming. We positioned it relative to specification-driven and requirements-driven design---including industry practice such as GitHub Spec Kit and Microsoft specification-driven development---systematic design frameworks, and AI-assisted design. We demonstrated the workflow on an inverted pendulum control design project using two platforms---an open-source SimpleFOC reaction wheel pendulum and a commercial Quanser Furuta pendulum---showing how the same specification-driven workflow and requirements can be satisfied by fundamentally different implementations, how domain decomposition was driven by literature references, how validation gates surfaced cost and feasibility constraints, and how user checkpoints ensured designer authority over decisions where literature was insufficient. \emph{Design-OS} extends the specification-driven multi-agent paradigm to physical engineering system design. We share a blank \emph{Design-OS} template and the full design-case artifacts in a public GitHub repository~\cite{DesignOS2026} to demonstrate generality and support replicability testing.
Future work will apply the workflow to further design cases, deepen integration with AI-assisted tools, evaluate in classroom settings, and conduct sandbox replicability testing across multiple AI models and independent users.

\balance
\renewcommand{\refname}{REFERENCES}
\bibliographystyle{config/asmems4}
\begin{mySmall}
\bibliography{References}
\end{mySmall}

\appendix
\xsection{Design-OS Commands}\label{app:commands}
\footnotesize
Table~\ref{tab:commands} lists the \emph{Design-OS} commands, organized by workflow stage. Each command is a structured Markdown prompt file that defines the stage's process steps, expected inputs and outputs, and validation checks. Commands are executed by invoking them as slash commands (e.g., \texttt{/plan-project}) in an IDE with appropriate AI model support (e.g., VS Code). They are model-agnostic by design. The full prompt files are available in the public repository~\cite{DesignOS2026} along with detailed explanations of the runtime requirements and individual commands.
\smallskip

\begin{table*}[t]
\def\arraystretch{1.0}
\centering
\caption{\emph{Design-OS} commands by stage.\label{tab:commands}}
\footnotesize
\begin{tabular}{@{}x{2.8cm}x{3.1cm}x{10.2cm}@{}}
\hline \hline
\textbf{Stage} & \textbf{Command} & \textbf{Purpose} \\
\hline
1: Concept Definition & \texttt{/plan-project} & Establish mission, roadmap, and tech stack through interactive dialogue \\
\hline
2: Literature Survey & \texttt{/literature-search} & Run landscape search and produce lit note, deep-research prompts, and scholar queries \\
  & \texttt{/focused-literature} & Targeted search driven by conceptual design gaps \\
  & \texttt{/verify-references} & Web-verify cited references and log corrections \\
\hline
3: Conceptual Design & \texttt{/conceptual-design} & Develop design objectives (DOs), methodologies, and high-level structure from literature \\
  & \texttt{/domain-decomposition} & Identify system domains and interfaces from literature references \\
\hline
Gate & \texttt{/validate-feasibility} & Gate check: cost, licensing, component availability, technical feasibility \\
\hline
4: Requirements Def. & \texttt{/spec} & Structure section-level requirements and specification layout (plan mode) \\
  & \texttt{/discover-standards} & Extract recurring design patterns into documented standards \\
  & \texttt{/index-standards} & Rebuild standards index for quick matching by \texttt{/inject-standards} \\
  & \texttt{/inject-standards} & Apply relevant standards to current work context \\
\hline
5: Design Definition & \texttt{/plan-implementation} & Derive task breakdown and dependency ordering from specification \\
  & \texttt{/generate-simulation} & Generate executable verification code from specification and model \\
\hline \hline
\end{tabular}
\end{table*}

\noindent The remainder of this appendix provides detailed definitions of each command. Each command file follows a common template: (1)~a preamble stating the command's role in the workflow, (2)~important guidelines that constrain the agent's behavior, (3)~a workflow validation block that checks prerequisites before execution begins, and (4)~a numbered process with 5--10 steps, each producing a named artifact.

\subsection{Stage 1: Concept Definition}

\paragraph{\texttt{/plan-project}.}
Establishes the foundational design context through interactive dialogue with the designer. The command walks through three areas: (1)~the design \emph{mission}---problem statement, stakeholders, success criteria, and constraints; (2)~a \emph{roadmap} defining the project phases, milestones, and deliverables; (3)~the \emph{tech stack}---tools, languages, simulation environments, and hardware platforms. The agent asks one question at a time, confirms each answer, and writes the results to three files (\texttt{mission.md}, \texttt{roadmap.md}, \texttt{tech-stack.md}) in the project folder. If existing project documents are found, the command reads them first and proposes updates rather than starting from scratch. The output forms the input context for all subsequent stages. Interactive checkpoints include: \emph{What system or design problem are we addressing?}, \emph{Who are the stakeholders or users?}, \emph{What constraints or success criteria matter?}, \emph{What are the main design phases or milestones?}, \emph{What comes first and what follows?}, \emph{What tools does the project use?}, and \emph{What formal methods or frameworks are you using?}

\subsection{Stage 2: Literature Survey}

\paragraph{\texttt{/literature-search}.}
Runs an initial landscape literature search aligned with the project mission and roadmap. The command reads the Stage~1 artifacts, then produces three outputs: (1)~\emph{deep-research prompts}---structured queries designed to be executed by AI research assistants (e.g., deep research tools) that request a single comprehensive report covering all relevant aspects of the design problem; (2)~\emph{scholar search queries}---keyword and Boolean queries formatted for academic search engines (Google Scholar, IEEE Xplore, etc.); (3)~a \emph{preliminary literature note} synthesizing what is known from the agent's training data, explicitly marking claims that require verification. The scope is deliberately preliminary: the command surfaces the landscape and identifies gaps rather than producing a final literature review. The designer is asked: \emph{What topics or domains should the literature search cover?}

\paragraph{\texttt{/focused-literature} (after Stage~3).}
Runs a targeted literature search driven by the conceptual design output. Unlike the initial landscape search, this command focuses on specific gaps, domain boundaries, or technical questions identified during conceptual design and domain decomposition. It requires the conceptual design artifacts as input, generates focused deep-research prompts and scholar queries scoped to the identified gaps, and produces a focused literature report. The command also supports an optional standards search when the conceptual design identifies regulatory or normative requirements. The designer is asked to \emph{confirm or adjust the focus areas} before the search proceeds.

\paragraph{\texttt{/verify-references} (after focused literature).}
Verifies all references cited in the preliminary and focused literature notes by web search. The command reads the literature artifacts, extracts every cited reference (author, title, year, venue), and checks each against web sources (publisher pages, Google Scholar, DOI resolvers). It produces an append-only \emph{corrections log} documenting: (1)~hallucinated references (no matching publication found), (2)~wrong-author attributions, (3)~incorrect years or venues, and (4)~minor errors (initials, title wording). The corrections log ensures that fabricated citations do not propagate into the specification. This command is fully automated and does not require user checkpoints.

\subsection{Stage 3: Conceptual Design}

\paragraph{\texttt{/conceptual-design}.}
Transforms the literature and project context into a first conceptual design. The command proceeds through: (1)~formulating \emph{design objectives} (DOs); (2)~identifying candidate \emph{methodologies} and approaches grounded in the literature; (3)~defining \emph{functions and needs} the system must satisfy; (4)~generating \emph{concepts} (candidate solutions or architectures); (5)~proposing a \emph{high-level structure} (e.g., paper outline, system architecture, or module breakdown). The command explicitly requires that design objectives and methodologies are fixed before concepts are generated, enforcing the principle that the problem framing precedes solution generation. Interactive checkpoints include: \emph{What are the primary design objectives?}, \emph{What formal methodologies will we use?}, \emph{What are the main functions and stakeholder needs?}, \emph{What solution concepts or platforms will we target?}, and a \emph{literature sufficiency check}.

\paragraph{\texttt{/domain-decomposition}.}
Identifies the constituent domains of the system---typically mechanical, electrical, and software for mechatronic systems---with every proposed domain boundary traceable to at least one literature reference from Stage~2. The command reads the literature and conceptual design artifacts, proposes domains with citations, identifies \emph{inter-domain interfaces} (e.g., the torque equation coupling electrical voltage to mechanical torque), and performs a \emph{gap assessment} that flags domain boundaries or interfaces where literature is insufficient. When gaps are found, the command surfaces a user checkpoint rather than fabricating justifications. Interactive checkpoints include: \emph{Are these domains correct? Should any be added, merged, or removed?}, \emph{What are the inter-domain interfaces?}, and a \emph{gap assessment} asking whether to flag gaps or provide information. The decomposition output feeds directly into Stage~4 (Requirements Definition), where requirements are organized by domain.

\subsection{Gate}

\paragraph{\texttt{/validate-feasibility}.}
Runs a feasibility gate check between major stages. The command can be invoked at any gate and assesses four dimensions: (1)~\emph{technical feasibility}---whether the proposed approach is realizable with the chosen architecture, components, and methods; (2)~\emph{cost assessment}---comparing platform and component costs (e.g., commercial vs.\ open-source hardware); (3)~\emph{licensing and IP}---evaluating software licensing constraints (proprietary vs.\ open-source), patent considerations, and data access; (4)~\emph{availability and procurement}---checking whether required components, platforms, or tools are accessible. The gate produces a feasibility report with candidate constraints that may become requirements in Stage~4. Interactive checkpoints include: \emph{Are any technical concerns blockers?}, \emph{Cost---estimates and budget constraints?}, and a \emph{go/no-go decision} on whether to proceed, iterate, or adjust scope.

\subsection{Stage 4: Requirements Definition}

\paragraph{\texttt{/spec}.}
Structures the specification by defining section-level requirements and the spec layout. This command must be run in plan mode (i.e., the agent plans the spec structure before writing it). It reads the conceptual design, domain decomposition, and literature artifacts, then produces: (1)~a \emph{shape} document defining the spec sections, their scope, and the requirement categories; (2)~a \emph{standards} document listing relevant standards and conventions discovered during the process; (3)~a \emph{references} document linking spec sections to their source literature. The spec folder structure is generated with placeholder files for requirements, interfaces, validation criteria, and traceability. Interactive checkpoints include: \emph{What are we designing?}, \emph{Are visuals needed?}, \emph{What reference implementations apply?}, and a \emph{consistency check} against existing project artifacts. The spec output serves as the scaffold for the elaborate-spec step (manual or command-driven), where detailed requirements with IDs, ``shall'' statements, priority, source, and verification method are filled in.

\paragraph{\texttt{/discover-standards}.}
Extracts recurring design patterns from the project into concise, documented standards stored in a dedicated \texttt{standards/} directory. The command analyzes the current project artifacts (code, specs, documentation), identifies patterns that should be codified (e.g., requirements format, naming conventions, system structure templates), and for each proposed standard asks the designer \emph{why} the pattern matters before writing it. Standards are indexed in a YAML file for discoverability. The designer is asked: \emph{Which areas should be documented as standards?} This command can be run at any point during the workflow but is most useful during or after Stage~4 (Requirements Definition) when patterns have emerged.

\paragraph{\texttt{/index-standards}.}
Rebuilds and maintains the standards index file (\texttt{standards/index.yml}). The index maps each standard to a brief description, enabling \texttt{/inject-standards} to suggest relevant standards without reading all files. The command scans all Markdown files in the \texttt{standards/} directory, compares them against the existing index, and for each new file proposes a one-sentence description for the designer to confirm. Deleted files are removed from the index. The command is also run automatically as the final step of \texttt{/discover-standards}. Interactive checkpoints include confirming or editing the proposed description for each new standard.

\paragraph{\texttt{/inject-standards}.}
Applies relevant standards from the \texttt{standards/} directory to the current work context. The command reads the standards index, analyzes the current task or artifact being worked on, matches applicable standards, and presents them to the designer with a recommendation. Standards can be injected automatically (e.g., when writing a new spec section, the requirements format standard is surfaced) or explicitly by the designer. This command ensures consistency across artifacts and stages without requiring the designer to manually consult the standards directory.

\subsection{Stage 5: Design Definition}

\paragraph{\texttt{/plan-implementation}.}
Turns an elaborated spec into a concrete implementation plan and task list. The command reads the spec folder (requirements, interfaces, validation criteria, traceability), detects the project context (e.g., control system, software module, mechatronic subsystem), and produces: (1)~an \emph{implementation plan} with ordered tasks, each tied to specific requirements and deliverables; (2)~a \emph{dependency graph} showing which tasks must precede others; (3)~a \emph{task list} (e.g., a checklist or todo list) for execution tracking. The designer confirms \emph{artifact structure, execution order, and deliverables}. The command ensures that every requirement in the spec has at least one corresponding implementation task.

\paragraph{\texttt{/generate-simulation}.}
Generates executable simulation code directly from the elaborated spec artifacts. The command reads the requirements document (specifically performance requirements with numerical targets), the system model (state-space matrices, transfer functions, or equations of motion), and the traceability table, then produces simulation code that: (1)~builds the system model from documented parameters, (2)~implements the control law or design under test, (3)~runs verification scenarios (e.g., step response, disturbance rejection), and (4)~checks each performance requirement against simulation results, reporting pass/fail with numerical values. The command chooses the simulation language based on the tech stack (e.g., Python or MATLAB) and ensures that the generated code is self-contained, reproducible, and directly traceable to the spec. This command is fully automated and does not require user checkpoints.



\end{document}